\documentclass[oneside,english]{mn2e}
\usepackage[T1]{fontenc}
\usepackage[latin9]{inputenc}
\usepackage{verbatim}
\usepackage{amsmath}
\usepackage{graphicx}
\usepackage{esint}

\makeatletter

\newcommand{\lyxdot}{.}

\makeatother

\usepackage{babel}

\begin{document}

\title[BH and Galactic Density Cusps I]{Black holes and Galactic density cusps -- I. Radial orbit cusps
and bulges}

\author[Le Delliou et al.]{M. Le Delliou$^{1}$\thanks{Email: Morgan.LeDelliou@uam.es (MLeD); henriksn@astro.queensu.ca (RNH);
joseph.macmillan@gmail.com (JDMcM)}, R.N. Henriksen$^{2}$\footnotemark[1] and J.D. MacMillan$^{3}$\footnotemark[1]\\
$^{1}$Instituto de Física Teórica UAM/CSIC, Facultad de Ciencias,
C-XI, Universidad Autónoma de Madrid\\
 Cantoblanco, 28049 Madrid SPAIN\\
$^{2}$Queen's University, Kingston, Ontario, Canada \\
$^{3}$Faculty of Science, University of Ontario Institute of Technology,
Oshawa, Ontario, Canada L1H 7K4}

\date{Send offprint requests to: MLeD\hfill{}Preprint: IFT-UAM/CSIC-09-26\\
Accepted 2010 December 18. Received 2010 December 13; in original
form 2010 August 12}
\maketitle
\begin{abstract}
In this paper, we study the distribution functions that arise naturally
during self-similar radial infall of collisionless matter. Such matter
may be thought of either as stars or as dark matter particles. If
a rigorous steady state is assumed, then the system is infinite and
is described by a universal distribution function given the self-similar
index. The steady logarithmic potential case is exceptional and yields
the familiar Gaussian for an infinite system with an inverse-square
density profile. We show subsequently that for time-dependent radial
self-similar infall, the logarithmic case is accurately described
by the Fridmann and Polyachenko distribution function. The system
in this case is finite but growing. We are able to embed a central
mass in the universal steady distribution only by iteration\textbf{,}
except in the case of massless particles. The iteration yields logarithmic
corrections to the massless particle case and requires a `renormalization'
of the central mass. A central spherical mass may be accurately embedded
in the Fridmann and Polyachenko growing distribution however. Some
speculation is given concerning the importance of radial collisionless
infall in actual galaxy formation.\end{abstract}

\pagerange{\pageref{firstpage}--\pageref{lastpage}}

\pubyear{2010}

\label{firstpage}

\section{Introduction}

The relation between the formation of black holes and of galaxies
has developed into a key astrophysical question, from the early papers
by Kormendy and Richstone \cite{KR1995}, to the more recent discoveries
by Magorrian et al. \cite{Ma98}, Ferrarese and Merritt \cite{FM2000},
and Gebhardt et al.\cite{Geb2000}. Recent papers \cite{Graham04,KB2009}
also establish a correlation between black hole mass and bulge luminosity
deficit. These papers establish a strong correlation between what
is essentially the black hole mass and the surrounding stellar bulge
mass (or velocity dispersion). The origin of this proportionality,
which extends well beyond the gravitational dominance of the black
hole, remains uncertain. But it is generally taken to imply a coeval
growth of the black hole and bulge.

Various proposals have been offered to explain the black hole mass-bulge
mass proportionality as a consequence of the AGN phase. There is as
yet no generally accepted scenario although a kind of `auto-levitation',
based on radiative feed-back from the accreting black hole to the
star-forming gas that in turns limit accretion, is plausible. In any
event there remains the question of the origin of the black hole seed
masses. In some galaxies at very high red shift the inferred black
hole masses are already of order $10^{9}~M_{\odot}$ \cite[e.g.]{Kurk2007}
after about one Ga of cosmic time. This may require frequent, extremely
luminous early events \cite[e.g.]{W09}, or it may suggest an alternate
growth mechanism.

The latter possibility is reinforced by the detection of a change
in the normalization of the black hole mass-bulge mass proportionality
in the sense of relatively larger black holes at high red shift \cite[e.g.]{Mai2007}.
As suggested in that paper it seems that the black holes may grow
first, independently of the bulge. The collisionless matter that we
invoke might be stars or it might be the dark matter itself.

Recently \cite{PFP2008} have studied the possible size of the dark
matter component in black hole masses. By assuming that the dimensional
or 'pseudo-phase-space density' \cite[ i.e. Henriksen 2006b, e.g.]{H2006}
is strictly constant they convert the relativistic accretion of the
dark matter into an adiabatic Bondi flow problem and obtain the resulting
accretion rate. Then by adopting the mass proportionality between
bulge and black hole and fitting boundary conditions from cosmological
halo simulations, they deduce that between 1\% and 10\% of the black
hole mass could be due to dark matter.

If we accept this result at face value, a seed mass of say $10^{6}M_{\odot}$
could have grown from dark matter. It would now be part of a super
massive black hole that subsequently grew in the AGN phase. Some seeds
may be primordial. As early as 1978, through fully general relativistic
numerical collapse calculations, \cite{BH1979} predicted primordial
black hole masses in the range $10^{2}$ to $10^{6}M_{\odot}$.

%
\begin{comment}
In this and subsequent papers we will attempt to embed a black hole
(or at least a central mass) in a distribution of particles that arises
naturally during the formation of the galactic core. We will predict
the consequent density cusp profile and that of the velocity dispersion
variation in the cusp. It is perhaps significant in the light of the
extensive orbital study of (\cite{V3CdZ08}) that we are most successful
in a time-dependent case. For these authors suggest that the central
mass may render the orbits chaotic and non-stationary. We find in
the steady systems that a series of iterations diverges at the centre.
\end{comment}
{}

Stellar density cusps surrounding black holes have been studied extensively
previously. Classic studies by \cite{P1972} and by \cite{BW76} dealt
with the problem of feeding the black hole from a filled loss cone
(nearly radial orbits). In addition to these diffusion studies, Young
\cite{Y1980} explored the cusps produced by the adiabatic growth
of a black hole in a pre-existing isothermal stellar environment.
This was extended by \cite{Q1995} and by \cite{MH2002} to more general
environments. In a cosmological context, Bertschinger \cite{Bertschinger}
also studied the growth of a central black hole by radial infall.
The conclusions were that the black hole induced cusps were never
flatter than $r^{-1.5}$ (the isothermal and cosmological case) and
that no black hole mass-bulge mass correlation was established \cite{MH2002}.
The latter conclusion has spurred the investigation of coeval dynamical
growth of the black hole and bulge in contrast to adiabatic growth
\cite{MH2003}. In this paper we discuss possible coeval growth due
to radial infall of stars and dark matter that both feed the black
hole and establish a collisionless cusp with a self-similar distribution
function.

{}

In this and subsequent papers \cite{HLeDMcM09a,HLeDMcM09b,HLeDMcM09c}
we will embed a black hole (or at least a central mass) in a distribution
of particles that arises naturally during the formation of the galactic
core. The `natural evolution' is taken to be that of self-similarity
since power-law behaviour is observed both in reality and in simulations.
In this fashion we introduce some uniqueness into the form of the
resulting distribution function that describes the collisionless matter
that surrounds the central mass. We will predict the consequent density
cusp profile and that of the velocity dispersion variation in the
cusp. 

In this first paper we confine ourselves to radial infall. The first
general result on radial infall was in fact given in \cite{FG84}
and the Bertschinger \cite{Bertschinger} result follows by putting
their parameter $\epsilon_{FG}=1/3$. However neither this work nor
that of Bertschinger attempted to infer closed forms for the equilibrium
distribution function. This was begun by Henriksen \& Widrow (1995)
(hereinafter \cite{HW95}). In \cite{H2006} it is pointed out that\textbf{
}$a=9/8$ ($\epsilon=3$) yields the Bertschinger solution in its
entirety, including the recently heralded power law of the proxy for
phase space density. The parameter $\epsilon=3\epsilon_{FG}$ and
it is related to the index called `$a$' below, both by simulations
and theory. 

We deduce in this paper two principal distribution functions that
arise naturally. One is steady and infinite and can not contain a
central mass exactly. However we can treat the central mass as a perturbation
to the gravitational field and iterate on the Boltzmann and Poisson
equations to find the flattest possible cusp in a steady infinite
system surrounding a central mass. There is a special logarithmic
case in this category that yields a Gaussian distribution function
and an inverse square law density cusp. Our derivation is new but
the result agrees with the form deduced in \cite{LB1967} based on
statistical mechanics.

Our second result is most important in that it represents a growing
cusp wherein a central mass may be embedded exactly. It is the Fridmann
and Polyachenko distribution function, which has not previously been
identified in this physical context. We derive it theoretically and
verify it with simulations. It also corresponds to a logarithmic potential
for which the self-similar index $a$ (see below) is unity. The compatibility
with a central mass allows us to give a growth rate for that mass.
It is perhaps significant in light of the extensive orbital study
of \cite{V3CdZ08} that we are most successful with our embedding
in a time dependent case. For these authors suggest that the central
mass may render the orbits chaotic and nonstationary.

We are aware that such a radial system is unstable to the radial orbit
instability (ROI) on small scales \cite[hereinafter for MacMillan et al. 2006]{MWH2006}.
However even fully cosmological simulations show an outer envelope
wherein the orbits are trending to be radial. Moreover isolated halos,
which also show statistical relaxation that is begging to be understood,
show quite radial orbits in the envelope \cite{MWH2006}. In such
cases the `central mass' may be a central `bulge' of dark matter that
forms rapidly. We speculate below that the growth of this bulge in
such an envelope may be described in terms of radial infall, and that
this infall may continue hierarchically to smaller scales after relaxation
by the ROI and clump-clump interactions \cite{MH2003}. Such interactions
would remove angular momentum from some particles in favour of others
and so create the continuing radial infall.

We begin the next section with the general formulation in spherical
symmetry. Subsequently in section \ref{sec:Radial-Orbit-Steady} we
discuss the various possible rigorously steady distribution functions
(DF from now on) for radial orbits. We show in section \ref{sec:The-Logarithmic-Case}
that the DF of Fridman and Polyachenko \cite[hereafter called FPDF]{FP1984}
describes a system of radial orbits that is growing self-similarly.
This is contrasted in the same section with an infinite steady system
for which the DF is Gaussian. After some discussion, we\textbf{ }give
our conclusions.

\section{Dynamical Equations in Infall Variables }

Following the formulation of \cite{H2006} in this section we transform
the collisionless Boltzmann and Poisson equations to `infall variables'.
We treat a spherically symmetric anisotropic system in the `Fujiwara'
form \cite[e.g.]{Fujiwara} namely

\begin{eqnarray}
 &  & \frac{\partial f}{\partial t}+v_{r}\frac{\partial f}{\partial r}+\left(\frac{j^{2}}{r^{3}}-\frac{\partial\Phi}{\partial r}\right)\frac{\partial f}{\partial v_{r}}=0,\label{eq:Boltzmann}\\
 &  & \frac{\partial}{\partial r}\left(r^{2}\frac{\partial\Phi}{\partial r}\right)=4\pi^{2}G\int f(r,v_{r},j^{2})dv_{r}dj^{2},\label{eq:Poisson}\end{eqnarray}

where $f$ is the phase-space mass density, $\Phi$ is the `mean'
field gravitational potential, $j^{2}$ is the square of the specific
angular momentum and other notation is more or less standard.

The `infall variables' are a system of variables and coordinates that
allows us either to readily take the self-similar limit or to retain
a memory of previous self-similar dynamical relaxation into a true
steady state. In this way we can remain `close' to self-similarity
just as the simulations appear to do. These coordinates \cite{H2006A,H2006}
allow the general expression of the Vlasov-Poisson set, but they also
contain a parameter ($a$) that reflects underlying self-similarity.
The self-similar limit is taken by assuming what we term `self-similar
virialisation', wherein the system is steady in these coordinates,
although it is not absolutely steady since mass is accumulating in
this mode. 

The transformation to infall variables has the form \cite[e.g.]{H2006}

\begin{align}
R & =r\, e^{-\alpha T/a},\hspace{2cm}Y=v_{r}e^{-(1/a-1)\alpha T},\nonumber \\
Z & =j^{2}e^{-(4/a-2)\alpha T},\hspace{0.95cm}e^{\alpha T}=\alpha t,\nonumber \\
P\left(R,Y,Z;T\right) & =e^{(3/a-1)\alpha T}\pi f\left(r,v_{r},j^{2};t\right),\label{eq:artrans}\\
\Psi\left(R;T\right) & =e^{-2(1/a-1)\alpha T}\Phi(r),\nonumber \\
\Theta\left(R;T\right) & =\rho(r,t)e^{2\alpha T}.\nonumber \end{align}

This transformation is inspired by the nature of self-similarity,
which can be understood as a scaling group wherein each quantity scales
according to its dimensions \cite{CH91}. The group parameter is the
logarithmic time $T$. The combinations of scaling constants (note
that $a\equiv\alpha/\delta$ see below) multiplying $\alpha T$ in
the exponential factor of each physical quantity reflect the dimensions
of that quantity. When the dependence on the parameter $T$ is retained
in the new variables, there is clearly no invariance along the scaling
group motion and so no self-similarity. This means that the passage
to the self-similar limit requires taking $\partial_{T}=0$ when acting
on the transformed variables. Thus the self-similar limit is a stationary
system in these variables, which is a state that we have described
elsewhere as `self-similar virialisation' \cite[i.e. Henriksen \& Widrow 1999]{HW99,LeD2001}.
The virial ratio $2K/|W|$ is a constant in this state (although greater
than one; $K$ is kinetic energy and $W$ is potential), but the system
is not steady in physical, i.e. untransformed, variables, since infall
continues. 

The single constant quantity $a$ is the constant that determines
the dynamical similarity, called the self-similar index. It is composed
of two separate scalings, $\alpha$ in time and $\delta$ in space,
in the form $a\equiv\alpha/\delta$. The dimensions of any mechanical
quantity can be expressed in the scaling space $\boldsymbol{a}\equiv(\alpha,\delta,\mu)$,
where $\mu$ is the mass scaling. The exponential factor of any physical
quantity $Q$ (which includes physical constants) is calculated as
$\boldsymbol{a}\cdot\boldsymbol{d}_{Q}$ where $\boldsymbol{d}_{Q}$
describes the quantity $Q$ in scaling space. Thus for the velocity
$\boldsymbol{d}_{v}=(-1,1,0)$, for the distribution function $\boldsymbol{d}_{f}=(3,-6,1)$
and for Newton's constant $\boldsymbol{d}_{G}=(-2,3,-1)$. In a gravitation
problem the scalings $\alpha$ and $\delta$ can express the scalings
of all physical quantities having mass, length and time dimensions.
This is because we require $G$ to be constant under the scaling motion
so that $\boldsymbol{a}\cdot\boldsymbol{d}_{G}=0$ and hence $\mu=3\delta-2\alpha$.
This changes $\boldsymbol{d}_{f}$ to $(1,-3)$ as used in equation
(\ref{eq:artrans}). This procedure yields the constants in the exponential
factors transforming all physical quantities in the equations (\ref{eq:artrans}).

We assume that time, radius, velocity and density are measured in
fiducial units $r_{o}/v_{o}$,$r_{o}$, $v_{o}$ and $\rho_{o}$ respectively.
The unit of the distribution function is $f_{o}$ and that of the
potential is $v_{o}^{2}$. We remove constants from the transformed
equations by taking \begin{equation}
f_{o}=\rho_{o}/v_{o}^{3},~~~~~~v_{o}^{2}=4\pi G\rho_{o}r_{o}^{2}.\label{eq:units}\end{equation}

These transformations convert equations (\ref{eq:Boltzmann}),(\ref{eq:Poisson})
to the respective forms

\begin{multline}
\frac{1}{\alpha}\partial_{T}P-\left(\frac{3}{a}-1\right)P+(\frac{Y}{\alpha}-\frac{R}{a})\partial_{R}P\\
-\left[\left(\frac{1}{a}-1\right)Y+\frac{1}{\alpha}\left(\frac{\partial\Psi}{\partial R}-\frac{Z}{R^{3}}\right)\right]\partial_{Y}P-\left(\frac{4}{a}-2\right)Z\partial_{Z}P=0\label{eq:SSBoltzmann}\end{multline}
 and \begin{equation}
\frac{1}{R^{2}}\frac{\partial}{\partial R}\left(R^{2}\frac{\partial\Psi}{\partial R}\right)=\Theta.\label{eq:SSPoisson}\end{equation}
 This integro-differential system is closed by \begin{equation}
\Theta=\frac{1}{R^{2}}\int~PdY~dZ.\label{eq:dmoment}\end{equation}

Until we enforce the self-similar limit ($\partial/\partial T=0$)
these equations remain completely general, because we have made a
continuous and invertible change of variables in equation (\ref{eq:artrans}).
The merit of the transformation at this stage is only that it puts
the expected asymptotic self-similar behaviour in the explicit exponential
factors, while relegating the declining time dependence leading to
this state to the transformed variables. These variables are strictly
independent of $T$ in the self-similar state. 

We will in this paper restrict ourselves to the filled loss cone limit
of radial infall \cite{HLeD2002}, although this is not the case in
subsequent papers of this series. This special case is certainly not
realistic where angular momentum becomes important, but it may have
application on large scales and in the subsequent evolution in regions
where angular momentum is transformed away either by bars or other
asymmetries. It may in any case be regarded as an introduction to
our methods.

To proceed we set\textbf{ }\[
P=F(R,Y;T)\delta(Z)=F(R,Y;T)\delta(j^{2})(e^{(4/a-2)\alpha T})\]

($\delta()$ is the Dirac delta, not the scaling delta) which changes
the scaling for the DF in equation (\ref{eq:artrans}) to \begin{equation}
\pi f=F(R,Y,;T)e^{(1/a-1)\alpha T}~\delta(j^{2}),\label{BoltzmannR}\end{equation}
 while other scalings remain unchanged.

The governing equations now become equation (\ref{eq:SSPoisson})
plus the Boltzmann equation for $F(R,Y;T)$ in the form

\begin{multline}
\frac{1}{\alpha}\partial_{T}F+\left(\frac{1}{a}-1\right)F+\left(\frac{Y}{\alpha}-\frac{R}{a}\right)\partial_{R}F\\
-\left[\left(\frac{1}{a}-1\right)Y+\frac{1}{\alpha}\frac{\partial\Psi}{\partial R}\right]\partial_{Y}F=0.\label{eq:SSBoltzmannR}\end{multline}
 Finally equation (\ref{eq:dmoment}) reduces to \begin{equation}
\Theta=\frac{1}{R^{2}}~\int~F~dY.\label{eq:dmomentR}\end{equation}

This completes the formalism that we will use to obtain the results
below.

\section{Steady Cusps and Bulges with Radial Orbits}

\label{sec:Radial-Orbit-Steady}

In this section, we find DFs both self-similar and steady which are
comprised of collisionless particles in radial orbits.\textbf{ }We
expect one mode of relaxation in collisionless cusps to be of the
`moderately violent' type satisfying, in terms of the particle energy
$E$ and mean field potential $\Phi$, the relation \begin{equation}
\frac{\mathrm{d}E}{\mathrm{d}t}=\frac{\partial\Phi}{\partial t}|_{r}.\label{eq:MVR}\end{equation}
 This includes phase-mixing. Another mode \cite{DKM2006} is furnished
by the presence of hierarchical sub-structure . The sub-structure
can interact in clump-clump interactions that can induce relaxation
on a coarse-grained scale \cite[i.e. Henriksen 2009]{H2009}.

However the temporal evolution of the system is difficult to follow
analytically even in the self-similar limit, so we normally look for
equilibria established by the evolution.This may be either a strictly
steady state in some appropriate coarse-grained description, or it
may be a self-similar virialised state.

In general one can not find a unique solution of the governing equations
that consistently reproduce infall onto a central mass. We find that
this is possible in one interesting case (Fridman and Polyachenko
DF) of accretion onto a point mass that arises naturally, but not
for a truly steady distribution around a point mass. One can allow
for the presence of a point mass in a rigorously steady distribution
by iterating about an equilibrium state that is determined initially
by the central mass. This allows the central mass and the environment
to be evolved together towards a new equilibrium, although normally
only a single loop is feasible (we give two loops below as a confirmation
of the continuing logarithmic behaviour). We proceed to derive these
two results in this section.

Using the characteristics of equation (\ref{eq:SSBoltzmannR}) plus
the total derivative \begin{equation}
\frac{\mathrm{d}\Psi}{\mathrm{d}s}=\frac{\partial\Psi}{\partial s}+\frac{\mathrm{d}R}{\mathrm{d}s}\partial_{R}\Psi\end{equation}
 where $\mathrm{d}s\equiv\alpha\mathrm{d}T$, one finds by a simple
manipulation that \begin{equation}
\frac{\mathrm{d}\left(\frac{Y^{2}}{2}+\Psi\right)}{\mathrm{d}s}=-2\left(\frac{1}{a}-1\right)\frac{Y^{2}}{2}-\frac{R}{a}\partial_{R}\Psi+\frac{\partial\Psi}{\partial s}.\label{eq:integral}\end{equation}
 In order for this last equation to yield the energy as an isolating
integral (i.e. characteristic constant), the sum of the last two terms
must give $-2(1/a-1)\Psi$. This is most simply effected by setting
$\partial\Psi/\partial s=0$ and $R\partial_{R}\Psi=p\Psi$, which
turns out to be a condition for both self-similarity and a true steady
state.%
\begin{comment}
,%
\footnote{The general solution to the sum condition is $\Psi=R^{p}G(r)$ where
the function $G(r)$ is arbitrary: however this leads only to the
general Jeans form.%
} wh
\end{comment}
{} Here \begin{equation}
p=2(1-a).\label{eq:mem1}\end{equation}
 so that $\Psi=\Psi_{o}R^{p}$ for some constant $\Psi_{o}$.

Hence, on setting $\mathcal{E}\equiv Y^{2}/2+\Psi$, we have from
equation (\ref{eq:integral}) \begin{equation}
\frac{\mathrm{d}\mathcal{E}}{\mathrm{d}s}=-2\left(\frac{1}{a}-1\right)\mathcal{E}.\label{eq:Echar}\end{equation}

This variation does render $E$ constant on characteristics (and therefore
in time) as one sees by integrating to the form $E_{o}\exp{(-2(1/a-1)s)}$,
and then by using the transformations (\ref{eq:artrans}) to find
$E=\mathcal{E}\exp{(2(1/a-1)\alpha T)}\equiv E_{o}$. An example of
such a state is a system of massless particles dominated by the potential
of a central point mass \textbf{$M_{*}$}, for which $a=3/2$ and
$p=-1$. The similarity index $a=3/2$ reflects the presence of the
Keplerian constant $GM_{*}$ whose vector $\boldsymbol{d}_{K}=(-2,3)$
and for which $\boldsymbol{a}\cdot\boldsymbol{d}_{K}=0$. We refer
to this example again below.

Equation (\ref{eq:SSBoltzmannR}) also yields along the characteristic
\begin{equation}
\frac{\mathrm{d}F}{\mathrm{d}s}=-\left(\frac{1}{a}-1\right)F,\label{DFR}\end{equation}
 so that with equation (\ref{eq:Echar}) \begin{equation}
F=\widetilde{F}(\kappa)|\mathcal{E}|^{1/2}.\label{eq:steadyF1}\end{equation}
The steady unscaled DF follows from this last equation and the transformations
(\ref{eq:artrans}) as ($a\ne1$) \begin{equation}
\pi f=\widetilde{F}(\kappa)|E|^{1/2}\delta(j^{2}).\label{eq:steadyF}\end{equation}

The quantity $\kappa$ in equation (\ref{eq:steadyF}) labels any
\emph{other} (besides $E$) possible characteristic constant, but
in general nothing other than $E$ is readily available. For this
reason we discuss the DF (\ref{eq:steadyF}) with $\widetilde{F}$
strictly constant.

In \cite{HW95} this DF was first given by assuming always a rigorous
steady state and was shown to yield the asymptotic particle distribution
found by Fillmore and Goldreich \cite{FG84}.Their result followed
from a direct integration of particle orbits in radial infall. The
example given in \cite{HW95} for comparison purposes corresponds
to the choice of our current index $a=18/17$. In general, $a=3\epsilon/2(\epsilon+1)$,
where the initial density profile is $\propto r^{-\epsilon}$. The
initial system is infinite for $\epsilon\le3$, which is $a\le9/8$.
In \cite{HW99} this DF was argued to be the natural state for steady
self-similarity. Thus the DF (\ref{eq:steadyF}) with $\widetilde{F}$
constant probably describes a steady system of self-similar radial
orbits characterized by the index $a$. For this reason together with
the numerical evidence discussed presently, we discuss the implications
of this DF here in more detail, bearing in mind the possibility of
embedding a central black hole or other spherical mass. %
\begin{figure*}
\includegraphics[clip,width=1\textwidth]{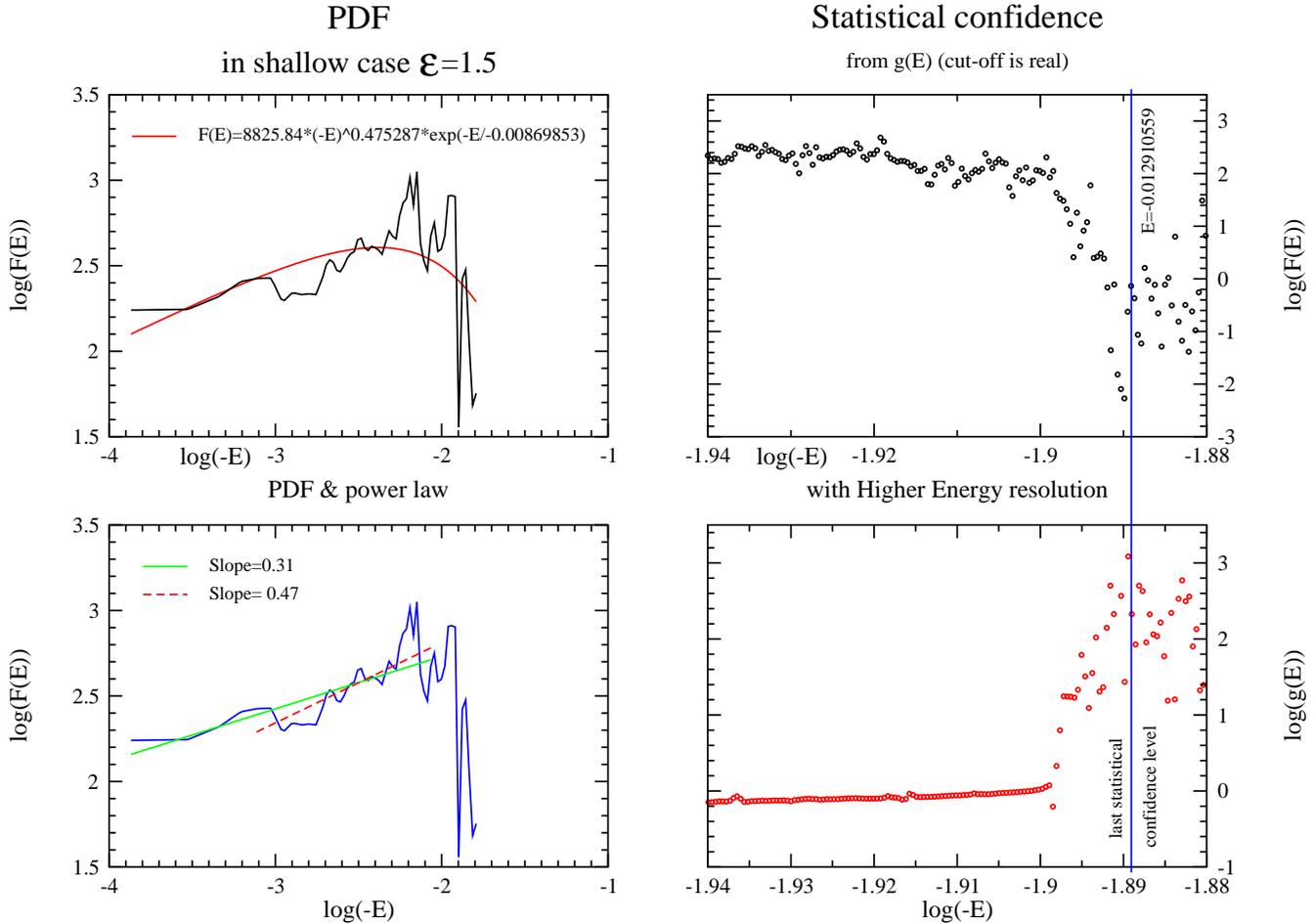}

\caption{\label{fig:-DF-shell}A shell code evaluation of the DF (Le Delliou
2001), evolved from a system with initial density $\rho\propto r^{-1.5}$.
The first fit is with a cut off power law ($F\propto E^{-p}e^{-E/E_{c}}$
with $p\simeq1/2$, $E_{c}\simeq-10^{-2}$; upper left panel), while
the second fit is just a power law (lower left), confirming (HW 1999).
The cut off is confirmed by higher resolution in the DF (upper right)
and in density of states $g(E)$ (lower right).}

\end{figure*}
In \cite{HW99} weak evidence was presented to show that the $|E|^{1/2}$
law did appear near the end of the infall for the most tightly bound
particles with negative energy. These would be closest to being described
as occupying a rigorous steady state. A similar result was found by
MacMillan \cite{MacM2006} for the most tightly bound particles even
while infall continued. The case was reinforced by additional calculation
of initially infinite systems by Le Delliou \cite{LeD2001} (see Fig.
\ref{fig:-DF-shell}). The figure shows another example of the Fillmore
\& Goldreich (1984) problem, wherein the DF (\ref{eq:steadyF}) is
a reasonable fit over most of the energy range. The cut-off is probably
numerical in this case since numerically the system is ultimately
finite.

However the DF (\ref{eq:steadyF}) does \emph{not} fit the complete
energy distribution found in high resolution simulations of radial
orbit \emph{growing} isolated halos \cite[e.g.]{MacM2006} in a state
of self-similar virialisation \cite{HW99}. In such a state, infall
continues. We reserve the explanation of this behaviour to the next
section where it involves the exceptional case wherein $a=1$.

Continuing for the moment with the discussion of equation (\ref{eq:steadyF})
we observe that an upper energy cut-off is required for finiteness
in the positive energy case ($a<1$ when the potential increases outwards),
while the cut-off is zero in the negative energy case ($a>1$ when
the potential decreases outwards). Moreover we note that one can add
an arbitrary constant $E_{o}$ to $E$ in this DF, which reflects
the arbitrary constant in the potential. For negative energy the DF
would appear as $\widetilde{F}(E_{o}-E)^{1/2}$ for $E<E_{o}<0$.
This was the type of fit used by MacMillan (2006) to fit his simulations.
The DF decreases to zero at $E_{o}$ and increases to negative energy.
A large positive constant $E_{o}$ and \textbf{$E<E_{o}$} so that
\textbf{$f\propto(E_{o}-E)^{1/2}$} would express a positive energy
cut-off at the value \textbf{$E_{o}$} and the DF would increase towards
zero energy.

The potential and density pair for these rigorously steady models
take the form%
\footnote{We have used our current notation when using the results of previous
papers, in which $\delta,X,S$ in \cite{HW99} are respectively $1/a,R\textrm{ and }\Theta R^{2}$.
In \cite{HW95} $\delta$ there becomes $1/(1-a)$ in current notation.%
} ($a\ne1$)

\begin{equation}
\Psi=\Psi_{o}R^{(2-2a)},~~~~~~\Theta=2(3-2a)(1-a)\Psi_{o}R^{-2a}.\label{eq:steadyH}\end{equation}

In a very recent submission Amorisco and Evans \cite{AE2010} prove
that the power-law relation between the galactic half-light radius
and the central velocity dispersion in dwarf ellipticals requires
a power law potential to be valid. This arises naturally here.

In the example of a dominant central point mass, inserting the index
$a=3/2$ in the above pair yields a point mass surrounded by massless
particles. The massless particles may be distributed in any manner,
but since\textbf{ $\boldsymbol{d}_{N}=(0,-3)$ }dimensional analysis
requires that the number density $N$ should vary as $N\propto N_{o}(R)e^{-3\delta T}=N_{o}(R)R^{3}/r^{3}$.
In a rigorously steady state this should be time independent so that
$N_{o}(R)\propto R^{-3}$ and hence \textbf{$N\propto r^{-3}$}. Such
a halo could exist outside any dominant mass as was discussed in \cite{HW99}.
Thus it might surround a central point mass or indeed be the diffuse
halo around a bulge containing most of the system mass. However this
is only a limiting behaviour and does not include the transition region.
This region interests us particularly in the context of central black
holes.

The direct density integral over the DF (\ref{eq:steadyF}) for negative
energies yields for $\rho$ \begin{equation}
\rho=\frac{\pi\widetilde{F}}{\sqrt{2}}\frac{|\Phi|}{r^{2}}.\label{eq:lindens}\end{equation}
 Since this is linear in the potential, one can readily include a
central mass by iteration. We may begin with a point mass potential
for $\Phi$ in the density (\ref{eq:lindens}), and then use the Poisson
equation to obtain a new potential in a form that is no longer self-similar.
This yields \begin{equation}
\Phi_{1}=-\frac{M_{\star}+C_{2}(1+\ln{r})}{r},\end{equation}
 whence follows a new density by (\ref{eq:lindens}). The constant
$M_{\star}$ would be the mass inside\textbf{ $\ln{r}=-1$ }(regarded
as a point mass) while\textbf{ $C_{2}=\left(\pi\widetilde{F}/\sqrt{2}\right)M_{\star}$}.
There is only a logarithmic modification to the $r^{-3}$ law at large
$r$ where the iteration should apply. In effect the iteration yields
a singular perturbation series at $r=0$ because of the diverging
potential and hence energy. Therefore we arbitrarily cut off the series
at small $r$ and `renormalize' the central mass to the mass inside
this cut-off radius. The next loop of the iteration gives\textbf{
\[
\Phi_{2}=-\frac{M_{*}(1-\frac{C_{2}^{2}}{2M_{*}^{2}})}{r}-C_{2}(1+\frac{2C_{2}}{M_{*}})\frac{(1+\ln{r})}{r}-\frac{\ln^{2}{r}}{r}\frac{C_{2}^{2}}{2M_{*}}\]
} where again we have `renormalised' at $\ln{r}=-1$. Putting this
back into equation (\ref{eq:lindens}) yields $\rho_{2}$ as $r^{-3}$
flattened only by logarithmic terms at large $r$ \cite{HW99} as
expected. The large scale $r^{-3}$ density profile does not fit the
bulge simulations inside the NFW \cite{NFW} scale length, but it
does describe the halo region outside a central bulge of mass $M_{\star}$
\cite{HW99}.

Since the density is linear in the potential we may also solve for
a self-consistent cusp having the DF (\ref{eq:steadyF}) by letting
the potential be determined by the Poisson equation. Working in transformed
variables we find \begin{equation}
\Psi=-AR^{p_{-}}-BR^{p_{+}},\label{eq:screenR2}\end{equation}
 where $A$, $B$ are arbitrary real constants $>0$ and \begin{equation}
p_{\pm}=-\frac{1}{2}\pm\sqrt{\frac{1}{4}-\frac{\pi}{\sqrt{2}}\widetilde{F}}.\label{eq:mem2}\end{equation}

By letting $\widetilde{F}\rightarrow0$ we see that $p_{-}$ is the
power that should be taken near the centre if we wish to create a
strong central mass concentration. It tends to $-1$ in this limit
while $p_{+}$ tends to zero. Hence we set $B=0$ in this limiting
domain. The potential then satisfies our basic condition (\ref{eq:mem1})
with a new self-similar index. This is given by $a=1-p_{-}/2$ according
to equation (\ref{eq:mem1}), that is explicitly \begin{equation}
a_{-}=\frac{5}{4}+\frac{1}{2}\sqrt{\frac{1}{4}-\frac{\pi}{\sqrt{2}}\widetilde{F}}.\label{eq:mem3}\end{equation}
Equation (\ref{eq:steadyH}) now gives the inner cusp density law
as \begin{equation}
\Theta=|p_{-}|(1+p_{-})|\Psi_{o}|R^{(-2+p_{-})}.)\label{eq:Rcuspscreen}\end{equation}
 This can not be flatter than $R^{-2.5}$, which appears only for
the ` maximum bulge' for which $\widetilde{F}=1/(2\pi\sqrt{2})$.
At large $r$ the term in $p_{+}$ dominates, and the behaviour tends
to $r^{-2}$ for $\widetilde{F}$ small.

In the context of dark matter simulations such a steady halo of radial
orbits could describe the region just beyond the NFW scale radius
(which we take to form the `bulge'), based on the density profile
alone. It is not stable in a strictly steady state according to the
usual Antonov criteria \cite[e.g.]{BT1987}, unless the energy is
negative. Consequently we do not expect it in central regions where
$a<1$ and the energy is positive (with a central zero: the potential
increases outward according to Eq.~\ref{eq:steadyH}). The radial
velocity dispersion is $\overline{v_{r}^{2}}=|\Phi|/2$. 

This concludes our study of the general, steady, spherically symmetric,
DF for power law distributions of radial orbits. The main justification
for the study is that it permits definite conclusions. %
\begin{comment}
The effect of an embedded mass is more problematic, but iteration
does suggest the $r^{-3}$ law to within logarithmic corrections.
\end{comment}
{} The rigorously steady DF (\ref{eq:steadyF}) is not permitted to
contain a black-hole or other centralised spherical mass without perturbation.
By iteration we show that it can only yield logarithmic corrections
to an inverse cube law. This suggests that a black hole will engender
time dependence in its surroundings.

The DF (\ref{eq:steadyF}) also permits a self-consistent exact solution
for the potential and density in the cusp (see e.g. equation \ref{eq:Rcuspscreen}).
However it fails to produce a sufficiently flat cusp of dark matter
or stars to explain observations of the Milky Way cusp of stars. At
large distance it yields (at flattest) an inverse square law density.

For these reasons we turn in the next section to a time dependent
case which is much more promising. It allows us to calculate the growth
of a central mass due to infall of collisionless matter on radial
orbits.

%
\begin{comment}
In the next section we turn to the exceptional case for which $a=1$.
We are able to place this case in the context of other discussions.
\end{comment}
{}

\section{The Logarithmic Case}

\label{sec:The-Logarithmic-Case}

The case $a=1$ is obviously special and, as it turns out, rather
important. We return to the equation (\ref{eq:integral}) and observe
that we also have an integral in the steady state if $\psi=\psi_{o}\ln{R}$.
For in that case the equation integrates to $\mathcal{E}+\psi_{o}\alpha T=\kappa$,
where $\kappa$ is constant on the characteristic. Equation (\ref{DFR})
then requires that in general $F=F(\kappa)$, so $F$ is also constant
on the characteristic. With $a=1$ we note from equation (\ref{eq:artrans})
that $\Psi\equiv\Phi$, $Y\equiv v_{r}$ and so $\mathcal{E}\equiv E$.
However $\kappa=v_{r}^{2}/2+\psi_{o}\ln{R}+\alpha T=v_{r}^{2}/2+\psi_{o}\ln{r}\equiv E$
and moreover $\pi f\equiv F$. Hence our conclusion is for the moment
only that \begin{equation}
\pi f=F(E),\end{equation}
 in this case. 

By taking the potential to be logarithmic, $\Psi=\Psi_{o}\ln{R}$,
we require by equation (\ref{eq:SSPoisson}) that $\Theta=\Psi_{o}/R^{2}$
and hence, by the appropriate members of the set (\ref{eq:artrans}),
that $\rho=\Psi_{o}/r^{2}$. But for consistency we must also have
\begin{equation}
\rho=\frac{\Psi_{o}}{r^{2}}=\frac{1}{r^{2}}\int~F(E)~\mathrm{d}v_{r}.\label{eq:consist}\end{equation}
 We must therefore find a DF $F(E)$ which satisfies this equation.
In effect, the integral over the particle velocities must be a constant
independent of the logarithmic potential. 

To find such a DF we convert the integral to an integral over energy
in the normal fashion and write our consistency condition as \begin{equation}
I\equiv\sqrt{2}\int~\frac{F(E)}{\sqrt{(E-\Phi)}}~dE=\Psi_{o}.\label{eq:consistintegral}\end{equation}

We might expect a power law form for $F(E)$ on general grounds, and
given this a brief experimentation shows that the most general form
for $F(E)$ in this case may be written as (we suppose negative energy
to ensure convergence and $E<E_{o}<0$ where $E_{o}$ is an arbitrary
constant energy) \begin{equation}
F(E)=\frac{K}{\sqrt{(E_{o}-E)}}\label{eq:FP1}\end{equation}
 This may also be inferred as the unique solution with a finite energy
range by recognizing that equation (\ref{eq:consistintegral}) is
really a simple form of Abel's integral equation (see e.g. \cite{BT1987}
in the first or finite form, p651) whose solution is equation (\ref{eq:FP1})
with \begin{equation}
K=\frac{\Psi_{o}}{\sqrt{2}\pi}.\label{eq:KPsi_o}\end{equation}
 This may be checked by direct evaluation of the integral to find
that $\rho=\Psi_{o}/r^{2}$ as required.

This result holds only where $E<0$ and hence where $r<r_{o}$, where
$r_{o}$ is an arbitrary scale. However since we have not actually
set a fixed scale in the problem (which would entail setting $\delta=0$),
we have implicitly implied that $r/r_{o}=R/R_{o}$. Thus we have assumed
that $r_{o}=R_{o}e^{\alpha T}=\alpha R_{o}t$. We may take $R_{o}$
constant (dimensionless) so that there is a residual time dependence
because the outer boundary expands according to $r_{o}=R_{o}e^{\alpha T}$.
This implies that the mass inside $R_{o}$ and indeed inside any fixed
$R$ is growing as $M=4\pi\Psi_{o}R\alpha t$. 

We have thus succeeded according to the above in deriving the Fridmann
and Polyachenko DF \cite{FP1984} as the unique result of time-dependent
radial accretion of a growing inverse-square density `bulge'. This
is one of our major conclusions. 

\begin{figure}
\includegraphics[width=1\columnwidth]{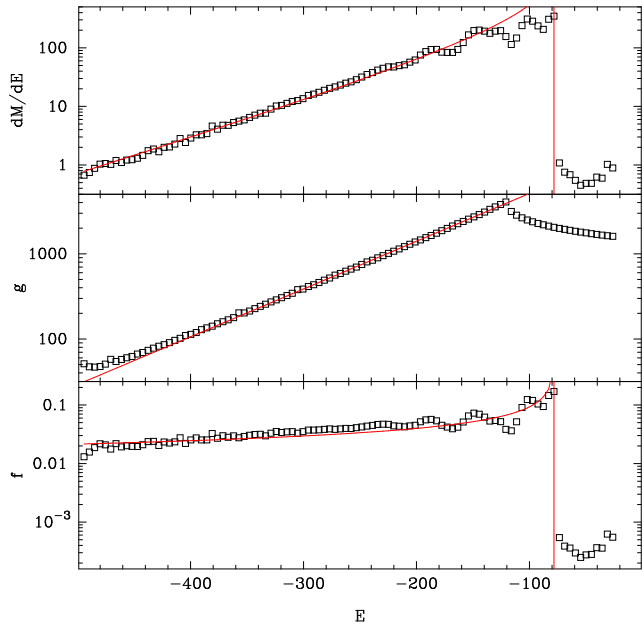}%{fig_5_8.eps}

\caption{\label{fig:fig_5_8} We show the Fridman and Polyachenko fit to the
mass distribution $dM/dE=f(E).g(E)$, density of states $g(E)$, and
the phase space distribution function $f(E)$. The figure is based
on the radial simulations of an isolated dark matter halo by (MacMillan
2006). The system is maintained in self-similar virialisation by steady
accretion. The fits use equation (\ref{eq:FPDF}) with $K=-E_{o}/(4\sqrt{2}\pi^{3})$
and $E_{o}\approx-80$ in machine units.}

\end{figure}

\begin{figure}
\includegraphics[width=1\columnwidth]{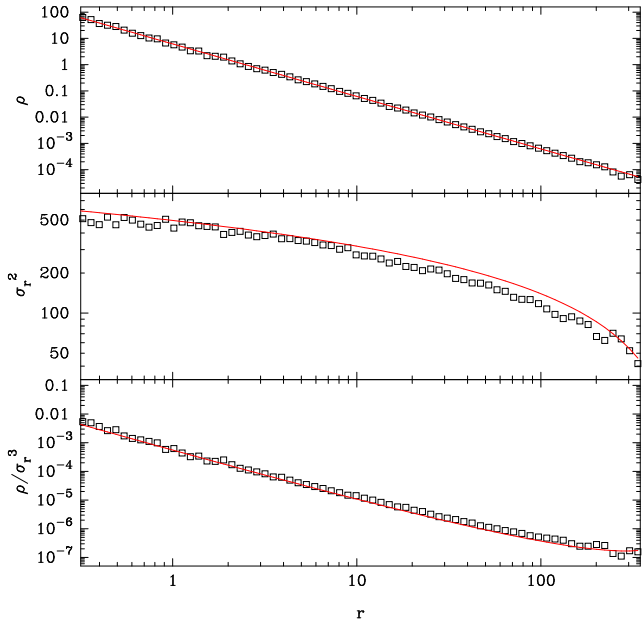}%{fig_5_11.eps}

\caption{\label{fig:fig_5_11} We show the the Fridman and Polyachenko fit
to the mass density, the velocity dispersion and the `pseudo-phase-space
density' for the same simulations by MacMillan.}

\end{figure}

A numerical measure of the DF in radial self-similar continuing infall
was made by MacMillan \cite{MacM2006}. Instead of the steady DF,
the DF of Fridman and Polyachenko \cite{FP1984} \textit{is found
to predict accurately all of the measured quantities} as in the accompanying
figures. These include an inverse square density law and a power law
pseudo-phase-space density of $\approx-1.5$, but there are logarithmic
corrections to the power law as can be calculated from the FP distribution
function. The pseudo-phase-space density power is flatter \cite[i.e. MacMillan, Widrow \& Henriksen 2006]{MWH2006}
than is generally found in full cosmological simulations. These results
are illustrated in figures (\ref{fig:fig_5_8}) and (\ref{fig:fig_5_11}).%

This DF \cite{FP1984} used to make the fits in figures (\ref{fig:fig_5_8})
and (\ref{fig:fig_5_11}) is \begin{equation}
f=\frac{K}{(-E+E_{o})^{1/2}}\delta(j^{2}),\label{eq:FPDF}\end{equation}
 for $E<E_{o}\le0$, and $r\le r_{f}$ (where $\Phi(r_{f})=E_{o}$)
and zero otherwise. This is just as we inferred above. The density
profile is $r^{-2}$ and the potential is logarithmic. \emph{The logarithmic
variation of the velocity dispersion together with the inverse square
density profile accounts for the pseudo density approximate power
law found in the simulations} \cite{MacM2006}.

The persistence of this DF is undoubtedly due to the strict proscription
of non-radial forces in the simulations. It is not linearly stable
by the Antonov criteria for $E<0$. When this proscription is relaxed
\cite{MacM2006} shows that the equilibrium Fridman and Polyachenko
DF is subject to the radial orbit instability. It may require continual
non-equilibrium excitation as provided by steady infall to be realized.

The unique feature of the distribution function (\ref{eq:FPDF}) is
that the density is independent of the potential. Hence one can simply
add a point mass potential to the logarithmic bulge value and the
density will remain $\Psi_{o}r^{-2}$. In the case of a true absorbing
central black hole one should only permit negative radial velocities
in the system. This means that $K$ in equation (\ref{eq:KPsi_o})
should be multiplied by a factor $2$. The velocity dispersion however
goes as $\overline{v_{r}^{2}}=|\Phi-E_{o}|$ and we may take $\Phi=-M_{\star}/r+\sqrt{2}\pi K\ln{r}+E_{o}$.
%
\begin{comment}
Normally the log term is negative since $r\le r_{f}$, and here $r_{f}=r_{o}$.
\end{comment}
{} MacMillan (2006) finds a good fit to this radial dispersion in his
simulations, but without a central mass.%

The actual growth of the black hole mass will be simply that of the
general self-similar mass growth as discussed above. That is \begin{equation}
M_{*}(t)=M_{*}(0)e^{\alpha T}=M_{*}(0)\alpha t.\label{eq:bhgrowth1}\end{equation}
 Its radius will be growing according to the same law. 

One can not however take seriously the growth of a black hole due
to radially infalling material from cosmological distances, since
the material there can not know the actual location of the black hole
and the radial infall is subject to the radial orbit instability.
However, as we will speculate in the discussion section, such growth
may apply to a hierarchy of `central' masses extending to ever smaller
scales. That is $M_{*}$ might be successively the bulge of a galaxy,
the core, the nucleus, and so on to the black hole. In each case there
must be a way of scattering orbits into the essentially radial loss
cone. Such scattering may be due in part to the formation of a bar
by the radial orbit instability itself, or to clump-clump interaction.

It is worth contrasting the above infall time-dependent DF with the
rigorously steady system of radial orbits in spherical symmetry. This
case was inadvertently omitted in the discussion by \cite{HW95} and
so we pause to present it in our general style as another example
of the method, similar to that above but in a strictly steady state. 

The main difference with the time dependent case is that that the
scaling motion must be taken in space rather than in time. This means
that we use the variable $R$, where $e^{\delta R}\equiv\delta r$
so that $dR/dr=e^{-\delta R}$. In addition we write \begin{equation}
f=F\delta(v_{\theta})\delta(v_{\phi}),\label{eq:steadyf}\end{equation}
 where $F$ satisfies {[}integrate the general steady CBE in spherical
symmetry\textbf{ }\cite{BT1987} over $v_{\theta}$ and $v_{\phi}${]}
\begin{equation}
v_{r}\frac{\partial F}{\partial r}-\frac{\partial\Phi}{\partial r}\frac{\partial F}{\partial v_{r}}+\frac{2Fv_{r}}{r}=0.\label{eq:steadyBoltzmann}\end{equation}
 In addition we have the Poisson equation as \begin{equation}
\frac{1}{r^{2}}\left[\frac{\mathrm{d}}{\mathrm{d}r}\left(r^{2}\frac{\mathrm{d}\Phi}{\mathrm{d}r}\right)\right]=\rho=\int~F~\mathrm{d}v_{r}.\label{eq:steadyPoisson}\end{equation}

The dimension vectors in our usual scaling space are $\boldsymbol{d}_{F}=(1,-4,1)$,
$\boldsymbol{d}_{v}=(-1,1,0)$,$\boldsymbol{d}_{\rho}=(0,-3,1)$ and
$\boldsymbol{d}_{\Phi}=(-2,2,0)$. Recalling that $\mu=3\delta-2\alpha$
the vector for $F$ becomes $(-1,-1)$, the vector for $\rho$ becomes
$(-2,0)$ while the others remain unchanged in the reduced $(\alpha,\delta)$
space. However we will only consider the case $a=1$ or $\alpha=\delta$
since the other cases were discussed in \cite{HW95} and reduce to
the DF (\ref{eq:steadyF}). When $\alpha=\delta$ the dimension vectors
reduce in delta dimension space to $(-2)$, $(-2)$, $(0)$ and $(0)$
respectively. Hence the equivalent of equation (\ref{eq:artrans})
is \begin{align}
\delta r & =e^{\delta R}, & Y & =v_{r},\nonumber \\
e^{-2\delta R}P\left(R,Y\right) & =F\left(r,v_{r}\right),\label{eq:sartrans}\\
\Psi\left(R\right) & =\Phi(r,t),\nonumber \\
e^{-2\delta R}\Theta\left(R\right) & =\rho(r).\nonumber \end{align}

If self-similarity is enforced normally, then in this case $\partial_{R}=0$
when it acts on the scaled variables. However this can not apply to
$\Psi(R)=\Phi$ since then equation (\ref{eq:steadyBoltzmann}) becomes
trivial. The answer lies in the realization that in this case the
potential is logarithmic in $r$, rather than being a power law. The
condition $\alpha=\delta$ requires that there be a constant $\Psi_{o}$
with dimensions of the potential per unit mass, just as in the time
dependent case. These conditions are satisfied here by setting \[
\Psi(R)=\Psi_{o}(\delta R)\equiv\Psi_{o}\ln{\delta r}=\Phi(r).\]
 Since only $\partial_{R}\Psi$ appears in the problem, the self-similar
requirement of independence of $R$ is maintained for $F$. 

A direct substitution of the transformation (\ref{eq:sartrans}) plus
the form of the potential into equation (\ref{eq:steadyBoltzmann})
reduces it to \[
Y\partial_{R}P-\frac{\mathrm{d}\Phi}{\mathrm{d}R}\partial_{Y}P=0\]

and hence after solving by characteristics \[
P=P(E)\]
 Consequently we find finally from $F=Pe^{-2\delta R}\equiv P/(\delta r)^{2}$
the %
\begin{comment}
eternally
\end{comment}
{} steady density in the %
\begin{comment}
Gaussian
\end{comment}
{} form

\begin{equation}
\rho=\frac{1}{\delta^{2}r^{2}}\int~P(E)~\mathrm{d}v_{r},\label{eq:GaussDF}\end{equation}

where \[
E\equiv\frac{v_{r}^{2}}{2}+\Phi(R).\]

From the Poisson equation (\ref{eq:steadyPoisson}) we obtain \[
\rho=\frac{\Psi_{o}}{r^{2}},\]
 and this must agree with the integral over $F$. %
\begin{comment}
The system is infinite and
\end{comment}
{} Letting both inward and outward going particles be present one finds
that \[
\rho=\frac{\sqrt{2}}{\delta^{2}r^{2}}\int~\frac{P(E)}{\sqrt{(E-\Phi)}}~\mathrm{d}E.\]
 %
\begin{comment}
Thus \[
K=\sqrt{\frac{\Psi_{o}}{\pi}},\]
 and there would be a factor two on the right if only inward (or less
likely) outward going particles were present. 

Such behaviour has been found previously as the steady, coarse-grained
limit both in spherical symmetry and in arbitrary symmetry \cite{H2004,LB1967}.
It is important to note that unlike the time-dependent result, this
eternally steady system is infinite in space and in mass. The Fridmann-Polyachenko
DF found for the time-dependent infall system, gives a finite although
growing mass. Hence it removes one objection to the Gaussian DF for
radial orbits. The contrast between the two solutions emphasizes the
difference between the completely relaxed DF in the steady state (\ref{eq:steadyF})
and the DF that is maintained by continuing infall (\ref{eq:FPDF}). 
\end{comment}
{}The problem is now reduced to the same situation that we had in the
time dependent case found earlier (\ref{eq:consist} and following).
In order to have a consistent (with the Poisson equation) inverse
square density law, only the steady Fridmann and Polyachenko DF may
be invoked, and that only for negative energies wherein $\Phi<E<E_{o}$.
Thus we derive the finite and steady Fridman and Polyachenko result
as quoted for example in \cite{BT1987}.

\section{Discussion and Conclusions}

In this paper we have presented unique distribution functions based
on the self-similar evolution of a gravitating system of radial orbits.
The rigorously steady state corresponds to an infinite system while
the time dependent system is finite and growing. Both of these systems
have been confirmed by numerical simulations, with the confirmation
of the growing mode being most dramatic.

Both systems have been derived coherently from the basic equations
by using a convenient formulation of self-similarity. We have considered
how these distributions are affected by the presence of a central
spherical mass, which may not necessarily be point-like. The steady
system is perturbed outside the central mass to an inverse cube law
with logarithmic corrections as we discuss further below. The growing
time-dependent case forms a perfect inverse-square density cusp that
can contain a central mass and is growing proportionally to the time
since formation. The material with which we are concerned is collisionless
and so may refer to dark matter in an early stage of formation or
to stars at a later stage. The restriction to radial orbits in this
paper is removed in subsequent papers, although both in theory and
in the numerical simulations a bias to radial orbits remains on the
larger scales.

For this latter reason we think that the results of this paper may
be relevant in reality for two reasons. In the first instance the
distribution functions may develop outside a rapidly formed large
central bulge wherein angular momentum has been important. This may
correspond to the NFW core radius, and our detailed results are compatible
with the density profile of the simulations outside this radius. The
second plausible scenario is that, on much smaller scales surrounding
an actual black hole, various instabilities may allow radial infall
of stars and dark matter onto the centre. This fits the time-dependent
infall model and allows the black hole to grow proportionally to the
time from the onset of the radial infall. We detail these results
in what follows.

In section (\ref{sec:Radial-Orbit-Steady}) on the steady state we
deduced the steady, spherically symmetric DF with radial orbits that
yields infinite systems with power law profiles. This was found previously
but we rederive it here as a limit of the time-dependent equations.
We have perturbed this DF by embedding a central mass. The DF is universal
for these systems ($f=K|E_{o}-E|^{1/2}$) but the potential-density
pair depends on the self-similar index $a$ which in turn depends
on dominant constants or boundary conditions. If it is a memory of
a cosmological fluctuation profile, then $a=3\epsilon/(2(\epsilon+1))$. 

This steady DF can not contain a central mass without being perturbed,
except for the Keplerian limit wherein $a=3/2$ and there is a mass-less
cusp with an inverse cube density profile. Iterating the equations
to find the perturbation produced by an embedded central mass predicts
a transition region in the halo of the mass that is an $r^{-3}$ profile
modified only by logarithmic corrections. It does not correspond to
the power law density of stars found close to the black hole in the
galactic centre\cite{G2009}, but may apply outside a bulge mass confined
to the NFW scale radius\textbf{.}

Since the density is linear in the potential we were also able to
find a solution \emph{that broke the self-similarity} by seeking non-power-law
solutions of Poisson's equation (pure power laws only exist in the
limits) with the density (\ref{eq:lindens}). The density profile
of the cusp can not be flatter than $r^{-2.5}$ near a central point
mass, but it may be as flat as $r^{-2}$ at large scales in the low
density limit. Such behaviour is too steep to explain the cusp observed
around the Sagittarius A{*} black hole. All possible descriptions
of a steady system of radial orbits are clearly excluded by this observation.
These descriptions are not excluded outside a central bulge however.

In the section concerning the logarithmic potential exception we showed
that it corresponded to continuing time-dependent infall. We found
both theoretically and numerically that the Fridmann and Polyachenko
DF is the distribution established by continuing radial infall. Remarkably,
it allows a growing central point mass (or indeed a bulge mass on
a larger scale) to be embedded in the infall self-consistently. %
\begin{comment}
We were able to contrast this result with a rigorously steady, infinite
system of radial orbits. This is described by a Gaussian DF that has
also been inferred in the past for fully relaxed, infinite systems
\textbf{\cite{LB1967}}.
\end{comment}
{} In both the steady and the time-dependent cases the density profile
is $r^{-2}$%
\begin{comment}
, but a black hole is not readily incorporated into the steady case
even by iteration
\end{comment}
{}. The time-dependent logarithmic case allows a consistent calculation
of the central mass growth rate according to $M_{*}\propto t$. Such
a growth rate from a surrounding envelope was found in \cite{MH2003}.
This simulation used a true N-body code that treated particle-particle
scattering, which scattering led eventually to radial infall of some
of the particles.

The growth rate of a central mass (i.e. a collisionless concentration,
not a true black hole) from a reservoir of radial orbits is zero if
there is a rigorous steady state. %
\begin{comment}
However in a state of self-similar virialisation we can expect them
to settle into a bulge as they become trapped by the increasing mass.
\end{comment}
{} The growth rate is not zero if the central mass is a black hole,
since then the outward bound radial orbits are suppressed. In that
case however the steady state is only an approximation except in an
infinite system. For a finite system the true timescale would be the
free-fall time of the bulk of the accreting mass.

One is inclined not to take these growth estimates seriously, since
they simply assume an endless supply of radial particles, and hence
an arbitrarily filled loss cone. In fact the radial alignment required
to hit a growing black hole from a few hundred parsecs is at least
one part in $10^{8}$ to $10^{10}$ depending on the mass of the black
hole! This suggests that instead \cite[see e.g.]{MH2003} the actual
growth involving radial orbits may be by way of a multi-stage process.
In the first stage, radial orbits accrete from the galactic halo to
form a bound spherical bulge of intermediate size, due to finite angular
momentum about the centre \cite{MH2003}. They are trapped there either
by the usual mechanism of self-similar infall as the potential increases
in time with increasing internal mass, or by dissipative interactions.
If there is substructure in the collisionless matter (e.g. stars and
dark matter clumps), then these are able to produce dissipational
collisions. Ultimately these collisions and tidal interactions can
lead to a more gradual growth of a more central mass \cite[e.g., in the Carnegie meeting]{MH2003}.

Recently a high resolution study \cite{Stetal2009} of sub-structure
in an isolated halo revealed that this sub-structure disappears in
the inner few parsecs. this coincides with the region where the halo
is becoming spherical and where the density power-law is flattening
to less that $1$. The interactions leading to the sub-structure disappearance
may well lead to relaxation.

In addition the radial orbit instability can lead to the development
of a bar \cite{MWH2006}. This bar can then transport angular momentum
away from the bulge by the ejection of particles. Such `interrupted
accretion' may repeat several times on the way to the actual central
object. The rapid accretion of a bulge is in fact the way in which
dark matter halos are thought to grow \cite{Zhao2003,Lu2006} initially.
This is then followed by a slower growth phase. The DF (\ref{eq:FPDF})
can be used to describe the environment of the central mass on each
scale of the interrupted cascade.

In this connection we refer to the work of Mutka \cite{Mutka09} on
gravitationally lensed galaxies with double images. He concludes that
there are two classes of density cusps with the larger sample (about
80\%) showing a logarithmic density slope of $\approx-1.95$ well
inside the NFW scale radius. The other 20\% show this slope as $\approx-1.45$.
These may be unresolved triple image lens and ,if so, the measured
value should be rejected.

Mutka's result is a measure of the total mass distribution rather
than just the dark matter. Perhaps we are seeing enhanced relaxation
in the mixture of stars and dark matter, that leads towards an isothermal
cusp, rather than the shallower cusps of the dark matter simulations.
It is significant that this inverse square slope is also frequently
found by direct dynamical modeling of galaxies \cite{vanderMarel09}.

However an inverse square slope is not restricted to a system of purely
radial orbits as the isotropic isothermal distribution shows. In a
subsequent paper we survey anisotropic distribution functions in spherical
spatial symmetry that also have a self-similar memory. Some of these
also provide an inverse square density profile.

The significance of the hierarchy of co-evolving structures is that
there will always be a mass correlation between them. Thus if the
mass concentration derives its ultimate mass $M_{\bullet}$ from a
halo of radius $r_{h}$, while $r_{s}$ encloses the mass that forms
the ultimate bulge mass $M_{s}$ then \begin{equation}
\frac{M_{\bullet}}{M_{s}}=\frac{r_{h}}{r_{s}}.\end{equation}
 This assumes the pure inverse square density law, which might in
fact have a logarithmic correction. In the subsequent paper, we shall
find a slightly more general correlation that involves the self-similar
memory. Taken at face value this simple relation gives $r_{h}/r_{s}\approx100$.

This paper comprises a systematic derivation of the distribution functions
that arise through self-similar evolution of gravitating systems of
radial orbits. In addition we have studied the effects of a central
mass on these distributions. We do not find density profile as flat
as those proposed in \cite{MS2006} and \cite{NM99} as a result of
density scouring by merging black holes. Thus our calculations probably
do not describe the cusp around Sagittarius A{*}, although this does
not exclude other applications. Subsequent papers in this series address
this problem by allowing anisotropy in velocity space.%
{}

\section{Acknowledgements}

RNH acknowledges the support of an operating grant from the Canadian
Natural Sciences and Research Council. The work of MLeD is supported
by CSIC (Spain) under the contract JAEDoc072, with partial support
from CICYT project FPA2006-05807, at the IFT, Universidad Autonoma
de Madrid, Spain%{Zel'dovich \& Podurets, 1965}
%\bibitem[Bertschinger, 1985]{Bertschinger85}Bertschinger, E., 1985, ApJS, 58, %39. 
%\bibitem[Del Popolo {\it et al}, 2000]{DelPopolo00}Del Popolo, A., Gambera, M.%, Rercami, Spedicato, E., 2000, A\&A, 353,
%427.

\label{lastpage}

\begin{thebibliography}{Zhao et al., 2003}
\bibitem[Amorisco\&Evans 2010]{AE2010} Amorisco N. C., Evans N. W.,
2010, preprint (arXiv:1009.1813A)

\bibitem[Bahcall \& Wolf 1976]{BW76} Bahcall J., Wolf R. A., 1976.
ApJ, 209, 214

\bibitem[1985]{Bertschinger} Bertschinger E., 1985, ApJS, 58, 39

\bibitem[Bicknell \& Henriksen 1979]{BH1979} Bicknell G. V., Henriksen
R. N., 1979, ApJ, 232,670

\bibitem[Binney \& Tremaine 1987]{BT1987} Binney J., Tremaine S.,
1987, \textit{\emph{Galactic Dynamics}}. Princeton Univ. Press, Princeton,
NJ

\bibitem[Carter \& Henriksen,1991]{CH91}Carter B., Henriksen R. N.,
1991, J. Math. Phys., 32, 2580

\bibitem[Diemand et al., 2006]{DKM2006}Diemand J., Kuhlen M., Madau
P., 2006, ApJ, 667,859

\bibitem[Ferrase \& Merritt 2000]{FM2000}Ferrarese L., Merritt D.,
2000, ApJ, 539, L9

\bibitem[Fillmore \& Goldreich, 1984]{FG84}Fillmore J. A., Goldreich
P., 1984, ApJ, 281, 1

\bibitem[Fridmann \& Polyachenko 1984]{FP1984}Fridman A. M., Polyachenko
V. L., 1984,\textit{\emph{ Physics of Gravitating Systems}}, Springer-Verlag,
New York (FPDF)

\bibitem[Fujiwara 1983]{Fujiwara}Fujiwara T., 1983, PASJ, 35, 547

\bibitem[Gebhardt et al., 2000]{Geb2000}Gebhardt K. et al., 2000,
ApJ, 539, L13

\bibitem[Gillessen et al., 2009]{G2009} Gillessen S., Eisenhauer
F., Trippe S., Alexander T., Genzel R., Martins F., Ott T., 2009,
ApJ, 692, 1075

\bibitem[H2004]{H2004} Henriksen R. N., 2004, MNRAS, 355, 1217

\bibitem[Henriksen 2006a]{H2006A} Henriksen R. N., 2006a, MNRAS,
366, 697

\bibitem[H2006]{H2006}Henriksen R. N., 2006b, ApJ, 653,894 %
\begin{comment}
\bibitem[H2007]{H2007} Henriksen, R.N., 2007, ApJ,671,1147.
\end{comment}
{}(H2006)

\bibitem[H2009]{H2009} Henriksen R. N., 2009, ApJ, 690, 102

\bibitem[HLeD 2002]{HLeD2002}Henriksen R. N., Le Delliou M., 2002,
MNRAS, 331, 423

\bibitem[HW95]{HW95}Henriksen R. N., Widrow L. M., 1995, MNRAS, 276,
679 (HW95)

\bibitem[HW 1999]{HW99}Henriksen R. N., Widrow L. M., 1999, MNRAS,
302, 321(HW1999)


%\bibitem[Jing \& Suto, 2000]{JingSuto}Jing, Y.P., Suto, Y., 2000, ApJ, 529, L6%9.


\bibitem[Graham 2004]{Graham04}Graham, A., 2004, ApJ, 613, L33

\bibitem[Kormendy \& Bender 2009]{KB2009}Kormendy J., Bender R.,
2009, ApJ, 691,L142

\bibitem[Kormendy \& Richstone 1995]{KR1995}Kormendy J., Richstone
D., 1995, ARA\&A, 33, 581

\bibitem[Kurk et al., 2007]{Kurk2007} Kurk J. D. et al., 2007, ApJ,
669, 32


%\bibitem[Kravtsov {\it et al} 1998]{Kravtsov98}Kravtsov, A.V., Klypin, A.A., %Bullock, J.S., Primack, J.R., 1998,
%ApJ, 502, 48.


\bibitem[Le Delliou 2001]{LeD2001}Le Delliou, M., 2001, PhD Thesis,
Queen's University, Kingston, Canada

\bibitem[I]{HLeDMcM09a}Le Delliou M., Henriksen R. N., MacMillan
J. D., 2010a, preprint (arXiv:0911.2232) (paper I)

\bibitem[II]{HLeDMcM09b}Le Delliou M., Henriksen R. N., MacMillan
J. D., 2010b, A\&A, 522, A28 (paper II)

\bibitem[III]{HLeDMcM09c}Le Delliou M., Henriksen R. N., MacMillan
J. D., 2011, A\&A, 526, A13 (paper III)

\bibitem[Lu et al. 2006]{Lu2006} Lu Y., Mo H. J., Katz N., Weinberg
M. D., 2006, MNRAS, 368, 1931

\bibitem[Lynden-Bell 1967]{LB1967} Lynden-Bell D., 1967, MNRAS, 36,
101

\bibitem[MacMillan 2006]{MacM2006}MacMillan J., 2006, PhD Thesis,
Queen's University at Kingston, Canada

\bibitem[MacMillan \& Henriksen 2002]{MH2002} MacMillan J. D., Henriksen
R. N., 2002, ApJ, 569, 83

\bibitem[MacMillan \& Henriksen 2003]{MH2003} MacMillan J. D., Henriksen
R. N., 2003, in Collin S., Combes F. and Shlosman I., eds., Meudon,
France, ASPC Proceedings, 290, 213

\bibitem[MWH 2006]{MWH2006} MacMillan J. D., Widrow L. M., Henriksen
R. N., 2006, ApJ, 653, 43 (MWH 2006)

\bibitem[Magorrian et al. 1998]{Ma98}Magorrian J. et al., 1998, AJ,
115, 2285

\bibitem[Maiolino et al. 2007]{Mai2007}Maiolino R. et al., 2007,
A\&A, 472, L33

\bibitem[Merritt \& Szell 2006]{MS2006} Merritt D., Szell A., 2006,
ApJ, 648, 890

\bibitem[Mutka 2009]{Mutka09} Mutka P., 2010, in Alimi J.-M., ed.,
Paris, France, AIP Conference Proceedings, 1241, 244. 

\bibitem[Nakano \& Makino 1999]{NM99}Nakano T., Makino M., 1999,
ApJ, 525, L77

\bibitem[Navarro, Frenk \& White 1996]{NFW}Navarro J. F., Frenk C.
S., White S. D. M., 1996, ApJ, 462, 5%63.


\bibitem[Peebles 1972]{P1972}Peebles P. J. E., 1972, Gen. Relativ.
Gravit., 3, 63

\bibitem[Peirani \& de Freitas Pacheco 2008]{PFP2008}Peirani S.,
de Freitas Pacheco J. A., 2008, Phys. Rev. D, 77, 064023


%\bibitem[Merrit {\it et al}, 1989]{MTJ}Merritt, D., Tremaine, S., Johnstone, D%., 1989, ApJ, 236, 829.


\bibitem[Quinlan et al 1995]{Q1995}Quinlan G. D., Hernquist L., Sigurdsson
S., 1995, ApJ, 440, 554


%\bibitem[Ryden, 1988]{Ryden88}Ryden, B.S., 1988, ApJ, 333, 78.


%\bibitem[Shapiro {\it et al}, 1986]{Shapiro86}Shapiro, S., Teukolsky, S.A., 19%86, ApJ, 307, 575.
%\bibitem[Sikivie {\it et al}, 1997]{Sikivie97}Sikivie, P., Tkachev, I.I., Wang%, Y., 1997, Phys. Rev. D, 56.
%\bibitem[Stiavelli \& Bertin, 1985]{StiavelliBertin}Stiavelli, M., Bertin, G., %1985, MNRAS, 217, 735.
%\bibitem[Stil, 1999]{Stil99}Stil, J., 1999, Doctoral Thesis, Leiden Observatory.
%\bibitem[Subramanian 2000]{Subra00}Subramanian, K., 2000, ApJ, in press
%
\begin{comment}
\bibitem[Tremaine 2005]{T2005}Tremaine, S., 2005, ApJ, 625,143.
\end{comment}
{}

\bibitem[Stadel et al 2009]{Stetal2009} Stadel J., Potter D., Moore
B., Diemand J., Madau P., Zemp M., Kuhlen M., Quilis V., 2009, MNRAS,
398, L21

\bibitem[ Van den Bosch et al. 2008]{V3CdZ08} Van den Bosch R. C.
E., van de Ven G., Verolme E. K., Cappellari M., de Zeeuw P. T., 2008,
MNRAS, 385, 647

\bibitem[van der Marel 2009]{vanderMarel09} van der Marel, R., 2009,
in S. Courteau ed., \textit{\emph{Unveiling the Mass}}. Queen's University,
Kingston, Ontario, Canada

\bibitem[Walter et al., 2009]{W09} Walter F., Riechers D., Cox P.,
Neri R., Carilli C., Bertoldi F., Weiss A., Maiolino R., 2009, Nature,
457, 699

\bibitem[Young 1980]{Y1980}Young P., 1980, ApJ, 242, 1232

\bibitem[Zhao et al., 2003]{Zhao2003} Zhao D. H., Mo H. J., Jing
Y. P., Börner G., 2003, MNRAS, 339. 12


%\bibitem[Zel'dovich \& Podurets, 1965]{Zel'dovich65}Zel'dovich, Ya.B;, Poduret%s, M.A., 1965, Soviet Astr.-A.J., 9, 742.


\end{thebibliography}
\end{document}